# BioNavi-NP: Biosynthesis Navigator for Natural Products

Shuangjia Zheng,[2,5,#] Tao Zeng,[1,#] Chengtao Li,[5] Binghong Chen,[3] Connor W. Coley,[4] Yuedong Yang[2,*] and Ruibo Wu[1,*]

[1] *School of Pharmaceutical Sciences, Sun Yat-sen University, Guangzhou 510006, P. R. China*

[2] *School of Data and Computer Science, Sun Yat-sen University, Guangzhou 510006, China*

[3] *College of Computing, Georgia Institute of Technology, Atlanta, GA, USA*

[4] *Department of Chemical Engineering, Massachusetts Institute of Technology, Cambridge, MA, USA.*

[5] *Galixir, Beijing, China*

*\*E-mail: wurb3@sysu.edu.cn, yangyd25@sysu.edu.cn*
*[#]These authors contribute equally to this work.*

**Abstract:** Nature, a synthetic master, creates more than 300,000 natural products (NPs) which are the major constituents of FDA-proved drugs owing to the vast chemical space of NPs. To date, there are fewer than 30,000 validated NPs compounds involved in about 33,000 known enzyme catalytic reactions, and even fewer biosynthetic pathways are known with complete cascade-connected enzyme catalysis. Therefore, it is valuable to make computer-aided bio-retrosynthesis predictions. Here, we develop BioNavi-NP, a navigable and user-friendly toolkit, which is capable of predicting the biosynthetic pathways for NPs and NP-like compounds through a novel (AND-OR Tree)-based planning algorithm, an enhanced molecular Transformer neural network, and a training set that combines general organic transformations and biosynthetic steps. Extensive evaluations reveal that BioNavi-NP generalizes well to identifying the reported biosynthetic pathways for 90% of test compounds and recovering the verified building blocks for 73%, significantly outperforming conventional rule-based approaches. Moreover, BioNavi-NP also shows an outstanding capacity of biologically-plausible pathways enumeration. In this sense, BioNavi-NP is a leading-edge toolkit to redesign complex biosynthetic pathways of natural products with applications to total or semi-synthesis and pathway elucidation or reconstruction.

**Introduction**

Natural products (NPs) are produced by organisms from all kingdoms in nature. To date, more than 300,000 NPs have been discovered and catalogued in libraries such as DNP[1] (Dictionary of Natural Products) and Super Natural II[2]. Remarkably, this vast chemical space of NPs is reachable from dozens of simple building blocks (also named as "chemo-bricks" in some literatures, as provided in Supplementary Figure 1). According to the classes of those building blocks, there are correspondingly four well-known biosynthetic pathways for major classes of NPs and their hybrids[3,4], including: (1) the AA/MA (acetic acid and malonic acid) pathway that produces fatty acids, phenols, and polyketides; (2) the MVA/MEP (mevalonic acid or methylerythritol phosphate) pathway that generates terpenoids and steroids; (3) the CA/SA (cinnamic acid or shikimic acid) pathway that yields flavonoids, phenylpropanoids, lignans, and coumarins; (4) the AAs (amino acids) pathway that constructs alkaloids. Unfortunately, as shown in Fig 1a, only about 33,000 enzymatic reactions have been characterized and confirmed, corresponding to fewer than 30,000 NPs serving as a substrate or product. That is, the complete biogenesis pathways including all intermediates are not established for most of the hundreds of thousands of known NPs. Accordingly, there are strong desires to reveal the biosynthetic pathways from essential building blocks to target NPs (namely the native NPs biogenesis) in the research fields of NPs.

It has been noted that NPs exhibit larger structural diversity and exist in a chemical space distinct from the fully synthetic molecules.[5] As a result, NPs play a significant role in the drug discovery and more than 60% FDA-proved small molecule drugs are NPs or their derivatives[6]. NPs are often the best option for seeking novel bioactive templates in human health as many of them are critical factors for regulating intrinsic biofunctions, especially in plants. However, obtaining commercially-relevant quantities of NPs is a major obstacle to their therapeutic translation. NPs are usually expressed in very low abundance in natural sources and thus conventional extraction approaches are inefficient and environmentally unfriendly in most cases. Meanwhile, many highly-valuable NPs have complicated structures and thus total synthesis is difficult and time-consuming. As famously exemplified by the heterologous biosynthesis of arteannuinic acid[7], biosynthesis and semi-synthesis of complex natural products has become a more popular and powerful strategy in the past decade with the theoretical advantages of lower cost and higher yield rate. Nevertheless, one of the common challenges for biosynthesis and semi-synthesis of NPs is the reconstruction of heterologous biosynthetic pathway from its native pathway. Computer-aided tools for the retro-biosynthesis analysis of target NPs are needed to increase adoption of these techniques, especially for the exploration of non-native biosynthetic pathways.

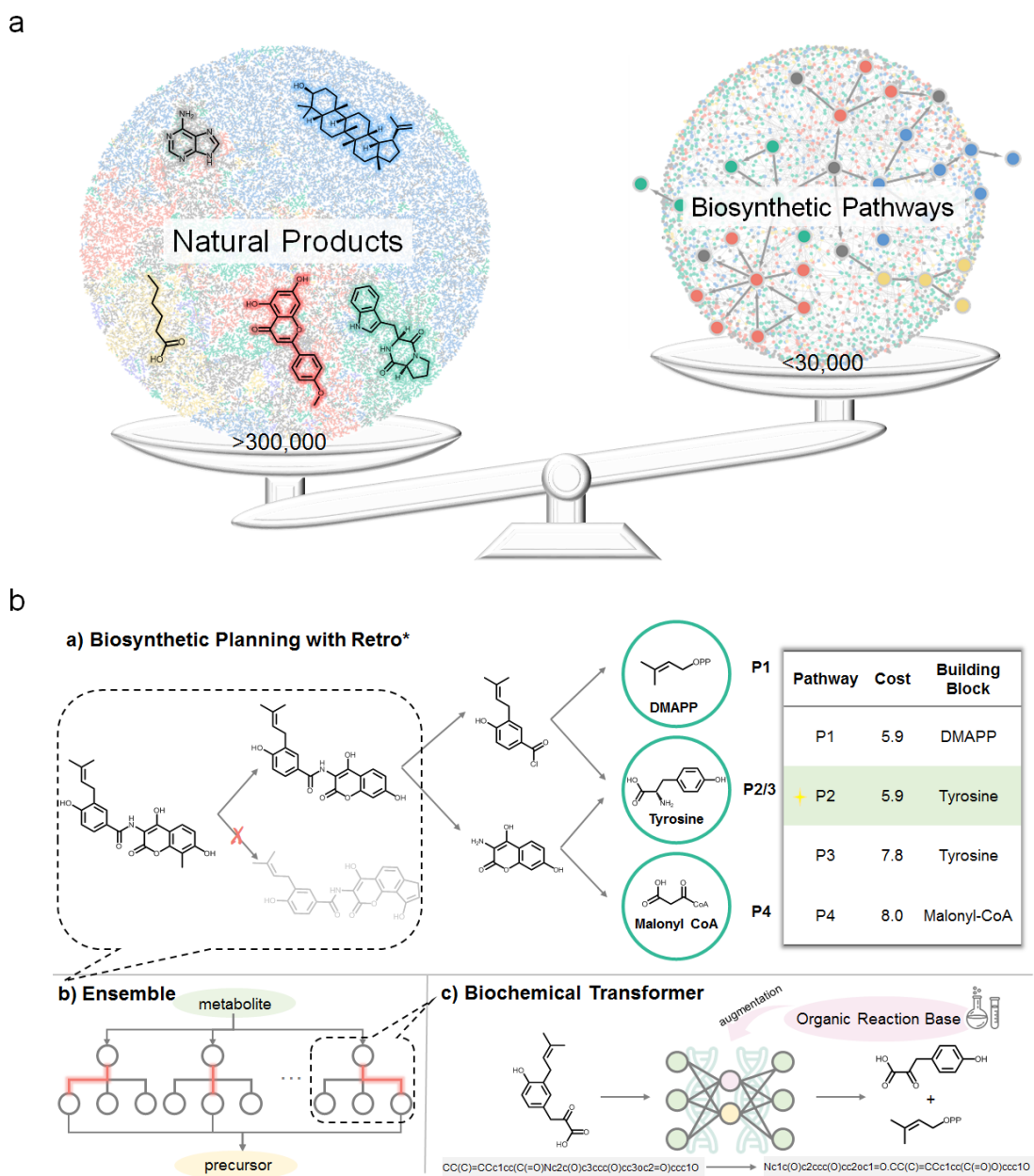

**Fig. 1 | a, The vast natural products and rare biosynthetic pathways reported to date. b, The protocol of BioNavi-NP to exploring biosynthetic pathways of target natural product.** The detailed models of Transformer and Retro* have been provided in Supplementary Figure 2.

To date, many efforts have been made in the field of retro-biosynthesis[8, 9], including the data-based similar pathway matching with known reactions[10, 11, 12], templated-based (reaction rule required) retro-biosynthesis prediction for short biocatalytic cascades[13, 14]. These tools will provide diverse options for the reconstruction of simple metabolites while they are rarely validated on the complex NPs. There are also works focused on the learning-based one-step reaction prediction[15, 16], retro-chemosynthetic planning for multi-step organic small molecule reaction[17, 18, 19], and computational expansion on the noscapine pathway to enrich benzylisoquinoline alkaloid derivatives[20]. However, these have not been applied to the retro-biosynthesis planning. Overall, there is yet no practical computational tool for retro-biosynthesis analysis, due to the multi-step and multi-pass features of biosynthetic pathways for most NPs.

Herein, we present BioNavi-NP as a new tool for proposing biosynthetic pathways to complex NPs from simple building blocks (Fig 1b). For a target NPs compound, its biosynthetic precursors are predicted by an ensemble of four Transformer neural networks that recursively propose one-step disconnections until plausible building blocks are identified. At each step, a deep learning-based multi-step planning strategy, Retro*[12], is used to manage the combinatorial number of options as the synthetic pathway's branches grow. Finally, the huge number of reaction pathways are ranked before the top-N pathways are visualized by an interactive website. The details for our computational protocols are provided in the Supplementary Methods Section 1-2.

## Results

**Single-step evaluation.** The multi-step retro-biosynthesis planning is based on the backward search performed through iterative applications of single-step retrosynthesis predictions. Therefore, it is critical to achieve a reliable prediction of single-step precursors at each step (rollout policy). As summarized in Table 1, the BioChem Transformer model directly trained on the biosynthesis dataset of 31710 reactions achieves top-1 and 10 accuracies of 10.6% and 27.8%, respectively. Training the model with 62370 organic reactions involving natural product-like compounds for augmentation (USPTO_NPL + BioChem) improves both accuracies to 17.2% and 48.2%, respectively. The >60% increase in accuracy indicates that the organic reaction expertise is helpful in predicting the biological synthesis process. Meanwhile, the pretrained model without fine-tuning does not make any correct predictions of biosynthetic precursors. This is in line with the fact that biosynthetic NPs and chemically synthetic compounds share some commonalities, but comprise two distinct structural spaces and distinct sets of reaction types, confirming that existing organic retrosynthesis tools cannot be directly used for biosynthesis prediction. Additionally, we found that correct handling of stereochemical information is essential for biosynthesis, as removal of chirality from the reaction SMILES decreases the top-10 accuracy from 27.8% to 16.3%. An additional 20% improvement in accuracy is obtained by an ensemble of four transformer models (four versions of USPTO_NPL + BioChem with different training steps), leading to top-1 and 10 accuracies of 21.7% and 60.6%, respectively. The performance is even comparable to those on predictions of organic retrosynthesis (the top-10 accuracy of ~60%)[21], suggesting the power of deep learning-based method for the planning of biosynthetic processes.

Table. 1 | **Performance of single-step models by different training strategies.**

| Training strategy* | top-N accuracy (%), n = | | | |
|---|---|---|---|---|
| | 1 | 3 | 5 | 10 |
| USPTO_NPL | 0 | 0 | 0 | 0 |
| BioChem (-chirality) | 7.6 | 11.1 | 13.9 | 16.3 |
| BioChem | 10.6 | 20.1 | 24.5 | 27.8 |
| USPTO_NPL + BioChem | 17.2 | 30.2 | 41.9 | 48.2 |
| USPTO_NPL + BioChem (x4) | 21.7 | 42.1 | 52.4 | 60.6 |

*USPTO_NPL includes 62370 NP-like organic reactions from the USPTO[22] database, BioChem includes 33710 biochemical reactions from MetaCyc[23], KEGG[24] and MetaNetX[25]. BioChem(-chirality) means no consideration of stereochemistry in the BioChem dataset. USPTO_NPL + BioChem (x4) means four different models obtained from training are ensembled to boost prediction performance. More introduction of the datasets can be found in the Supplementary Methods Section 3 and Supplementary Figure 3.

Table. 2 | **Comparison of performance among different models for test set.**

| Methods | Success rate | Hit rate of building block | Hit rate of pathway | Longest length | Ave. solution[a] | Time (h)[b] |
|---|---|---|---|---|---|---|
| BioNavi-NP (MCTS) | 128(34.8%) | 60 (16.3%) | 7(1.9%) | 3 | 1 | 92 |
| RetroPathRL | 194(52.7%) | 18(4.8%) | 14(3.8%) | 3 | 2.82 | **2** |
| RetroPathRL_UDB | 40(10.8%) | 19(5.1%) | 15(4.1%) | 3 | 2.82 | 3 |
| BioNavi-NP | **332(90.2%)** | 206(56.0%) | 91(24.7%) | **6** | **4.9** | 18 |
| BioNavi-NP_UDB | 275(74.7%) | **266(72.8%)** | **96(26.1%)** | **6** | **4.9** | 28 |

[a]Denotes the average number of pathways found, only the top-1 result is supported by MCTS variant, while for RetroPathRL, it outputs all pathway it can find. The output option for Retro* is set as top-5. [b]It is about 4-times computational cost for outputting top-10 in comparison to top-5, that is, the time consuming of BioNavi-NP (if only requesting the top-3) is comparable to RetroPathRL (the average number of pathways returned by RetroPathRL is close to 3). All data are obtained under the limit of 500 one-step calls. The test set includes 368 target NPs and more details are provide in the Supplementary Methods Section 3 and Supplementary Figure 3. Since the final goal of BioNavi-NP is aiming to develop a navigator of biosynthetic pathways network instead of a referee to pick up the best one, a comprehensive view of all outputs (here top-5 for comparison) is more important than find the so-called best one (top-1).

**Multi-step planning.** Based on the single-step biosynthesis prediction, we further explored the biosynthetic reaction space of NPs through multi-step bioreaction networks (internal cases, see Supplementary Methods Section 3). The comparison of chemical space for training set, internal cases and external cases is provided in the Supplementary Figure 4. As summarized in Table 2, we compared different methods in terms of their ability to predict biosynthesis routes that terminate in allowable starting materials ("success rate" or "solution rate"), to correctly find the reported pathway exactly as it appears in the knowledge base ("hit rate of pathway"), and to correctly recover the building blocks used in the known synthetic pathway ("hit rate of building block"). We note that this third metric is newly introduced in this work as the "ground truth" pathway could be not unique, there are multiple pathways between a natural product and its building blocks in many cases. For a candidate pathway linking a target molecule to building blocks, even if it is not the "ground truth" one, it could also be a meaningful non-native biosynthetic pathway or be an inspiration for pathway reconstruction and design.

BioNavi-NP outputs potential biosynthetic pathways for 332 out of 368 target NPs (90.2% success rate), a significant improvement over the available retro-biosynthesis pathway prediction tool RetroPathRL[13] (52.7%). By default, 40 building blocks (Supplementary Figure 1) are all used and the pathway search stops once it meets one of the building blocks. Regarding the hit rates of building blocks and pathway, BioNavi-NP also remarkably outperforms the RetroPathRL, with the hit rate of 56.0% and 24.7%, respectively. BioNavi-NP (MCTS), which substitutes Retro* with a Monte Carlo Tree Search (MCTS)[26], achieves a substantially lower success rate of 34.8%, indicating the effectiveness of Retro* strategy[18] for multi-step branching routes prediction. Considering that the building blocks of target NPs are well-known or user-desired in practical use, thus we also utilize the "ground truth" as the user-defined building blocks (UDB) for prediction, namely BioNavi-NP_UDB shown in Table 2. It is not surprising that both the hit rate of building blocks (72.8%) and pathways (26.1%) increased while the success rate (74.7%) decreased by pre-assigning the building blocks (only used the "ground truth" building block as the stopping criteria); the same trend is observed between RetroPathRL and RetroPathRL_UDB (Table 2).

To further evaluate the generalizability of BioNavi-NP, we collected additional 20 unseen NPs compounds (external cases) from recent publications that do not appear in the training and internal test sets and predict their plausible biosynthetic pathways. More than one candidate pathway was successfully found for each of the 20 cases (100% success rate), and the building blocks were correctly identified for 15 cases according to domain expertise (five cases are shown in Supplementary Figure 5). This suggests that BioNavi-NP using Retro* can be an effective tool for retro-biosynthesis analysis in terms of its ability to find reasonable building blocks and enumerating hypothetical biosynthetic pathways of target NPs. We note that the hit rate of pathway and building blocks might be increased by generating more pathways (at an increased computational cost) or further restricting allowable building blocks based on domain expertise.

To guide the reconstruction of biosynthesis pathway, it is more important to explore alternative pathways for target NPs. Therefore, in addition to the abovementioned hit rate of reported pathways, we also count the average number of predicted pathways as a direct indicator of the tool's use in idea generation. When selecting the output option as top-5 (see Table 2), BioNavi-NP outputs an average of 4.9, significantly higher than the 2.82 by RetroPathRL. The exploration ability was further confirmed by the longest length (6) of the pathways predicted by BioNavi-NP, in comparison to the three by RetroPathRL and BioNavi-NP (MCTS), indicating the ability of our model to more effectively address complex NPs.

In addition, we compared the exploration abilities and hit rates over different NPs families to understand the model's performance across different chemical spaces. For the five categories of natural products in the test set (including AA/MA, AAs, CA/SA, MVA/MEP and Others, as shown in Fig 2), BioNavi-NP achieved the highest hit rate of reported pathway (54.1%) and building block (94.6%) for the AA-MA category, while the lowest rate is only 19% for the prediction of the AAs biosynthetic pathways. This is likely attributable to the high similarity between some simple alkaloids and amino acids building blocks, as illustrated by the ellipse Fig 2b.

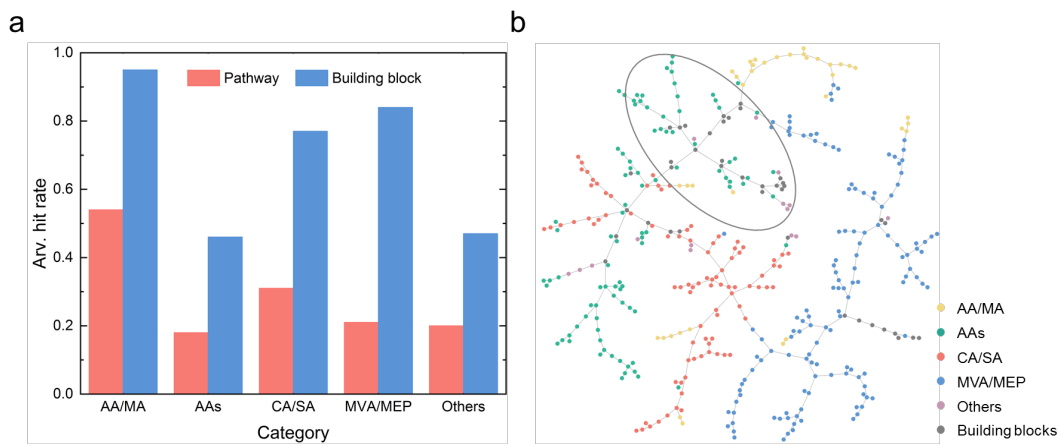

**Fig. 2 | a, The BioNavi-NP's performance within each NP category. b, The chemical space of compounds in each NPs category and building blocks.** The clustering and visualization of chemical space was realized by TMAP[27] using structural molecular fingerprints[28]. AA/MA: the acetic acid and malonic acid pathway, MVA/MEP: mevalonic acid or methylerythritol phosphate pathway, CA/SA: cinnamic acid or shikimic acid pathway, AAs: amino acids pathway.

**Web deployment and case study.** BioNavi-NP was implemented in a webserver for convenient redesign of biosynthetic pathways. As shown in Fig. 3, like a widely-used CASP tool ASKCOS[29] did, only the structure of target molecule is strictly required, but the default settings and building block list can be modified as needed. Predicted biosynthetic pathways are displayed in an interactive network, in which the target molecule can be traced back to the potential building blocks through several pathways along with the predicted cost of each step. To illustrate use of the webserver, we selected the sterhirsutin J and glutarate for the case studies. The server returns 6 and 10 candidate pathways, respectively, a subset of which are shown in Fig. 3b.

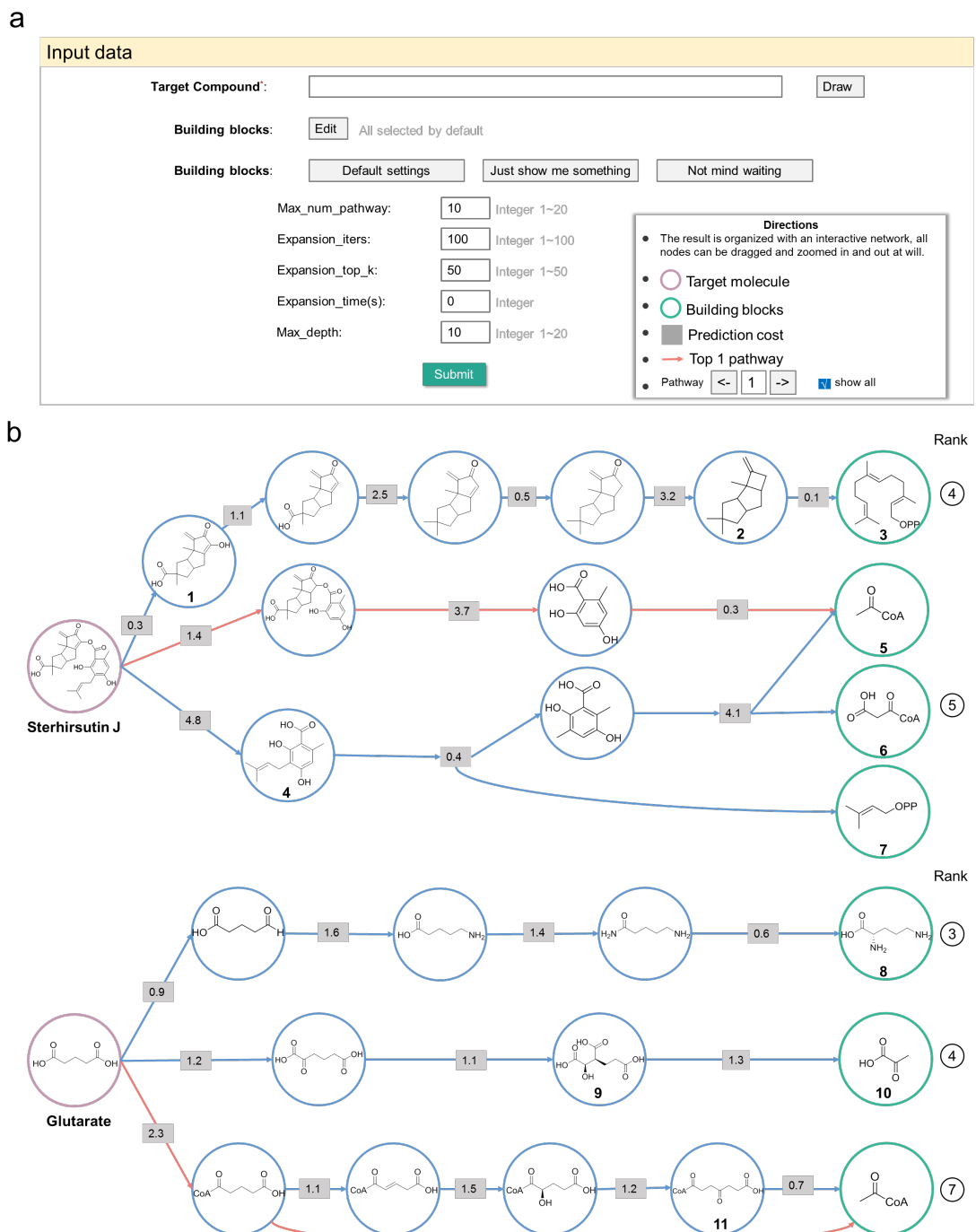

**Fig. 3 | a, The input interface and result legend panel of BioNavi-NP Web server. b, The selected output pathways of two typical examples (sterhirsutin J and glutarate) predicted by BioNavi-NP Web server.** The total number of the output pathways are determined by several options especially the "max number of pathways" (here default set is top-10), these default settings of several key options are given to balance the computational cost and accuracy. For clarity of the graphic illustration, the outputs are redrawn to be clear (the raw output found in Supplementary Figure 6-25) and part of candidate pathways with the order of ranks attached at the end, and the top 1 pathway is colored in red. There is also an option for which pathway to highlight or show. The cost of each reaction step is reflected by the confidence score (smaller prediction cost means higher reaction probability) and the total cost is the sum of all steps scores along the pathway.

Sterhirsutin J is a sesquiterpenoid derivative firstly isolated from the culture of *Stereum hirsutum* that has shown cytotoxicity against K562 and HCT116 cell lines.[30] As shown in Fig 4b, sterhirsutin J were decomposed into a hirsutane-type sesquiterpene (intermediate **1**) and colletorin D acid (intermediate **4**) to lead the fourth and fifth candidate pathways, respectively, where the fourth candidate pathway is ultimately traced back to building block farnesyl diphosphate (**3**), while the fifth candidate way is a hybrid biosynthesis style originated from the acyl CoA (**5**), malonyl CoA (**6**), and dimethylallyl diphosphate (**7**). Both of the fourth and fifth candidates are confirmed biosynthetic pathway according to the previous studies.[31, 32] This case shows us that BioNavi-NP is capable of dealing with complex structures including those derived from hybrid pathway and tracing them back to essential building blocks.

Glutarate (also called 1,5-pentanedioic acid) is an important raw material of organic chemical industry while biobased production of glutarate suffer from low titers.[33] Herein using BioNavi-NP, the predicted third candidate pathway belongs to one of the lysine (**8**) degradation pathways[34], and the seventh one is highly similar to a experimentally reconstructed pathway for glutaconate production[35], both of which are already existed in our training set, so it is not surprising that BioNavi-NP can predicted those two pathways which have been constructed in *E. coli* for glutarate production[36, 37]. The fourth candidate is not contained in our training set and it takes a successive decarboxylation strategy from homoisocitrate (**9**), which is predicted to be originating from the basic building block pyruvate (**10**). Interestingly, Wang et al.[38] recently established a novel glutarate biosynthetic pathway from α-ketoglutarate by incorporation of a "+1" carbon chain extension and α-keto acid decarboxylation, which covers the fourth candidate pathway predicted herein. This case study is a typical example to show the capability of BioNavi-NP in exploring the novel biosynthetic pathways beyond the known biogenesis, thus it is very useful for biosynthetic pathways redesign.

**Discussions**

In contrast to earlier works, BioNavi-NP is a deep learning-based end-to-end model that was constructed based on biosynthetic and organic reaction data without needs of cumbersome extraction of reaction templates or biased expert systems. This Transformer-based, template-free method presents a high success rate at generating routes to NPs that partially recapitulate known pathways and known building blocks, particularly in comparison to other methods at comparable computational costs (Table 2). Nevertheless, there are several opportunities to improve BioNavi-NP in future work:

(**a**) Intermediates may be incorrect or missing in some predictions. For example, the fourth candidate pathway of sterhirsutin J involves an unreasonable intermediate **2** which deviates from the well-known C5 rule of terpenoids. Another example is that intermediates between **9** and **10** are missing according to ref. [38]. This reminds us that BioNavi-NP is like a good navigator instead of a scorer, and thus we should comprehensively view the whole pathways network if computational cost is affordable. As exemplified by sterhirsutin J, there are missing intermediates/building blocks if considering the fifth and sixth candidates separately, but the suggestion is reasonable if integrating the two candidates together. In other words, to achieve the optimal navigation in the complicated biosynthetic pathways network of NPs, it is better to consider the whole candidate pathways network first; integration or correction may be required for rational design of high-efficient biosynthetic pathway.

(**b**) The scoring function does not provide the ideal ranking of potential pathways. As shown in Fig 3, BioNavi-NP favors the shortest pathway as the top-1 since the lower prediction cost is given by the intrinsic scoring function of Retro*. Generally, the scoring values do not have a quantitative meaning and the order of candidates does not reflect their chemical feasibility. Nevertheless, it might be useful for finding shortcuts in some cases. For example (Fig 3b), the predicted top-1 pathway of glutarate is rational based on chemical intuition. An effective way to use our BioNavi-NP is to overview the whole network first and then select the optimal candidate by experience and chemical logic. When BioNavi-NP find an unreasonable pathway in the first prediction, it is worth a second prediction to using a well-validated or user-confident intermediate as the target molecule.

(**c**) Specific enzymes are not predicted in the current version of BioNavi-NP. To date, detailed enzyme information in databases is not sufficient for meaningful learning on most known enzymatic reactions. And just like the trade-off between generalization and specificity for template-based organic reaction planning[39], the promiscuity, fidelity and diversity of enzymatic catalysis make it challenging to enzyme assignment for a given bioreaction. The biosynthetic pathways designed by BioNavi-NP just handle the first step to map out the biosynthesis networks of NPs, additional tools such as Selenzyme[40] might be complementary, and the interplay between theoretical prediction and experimental validation is necessary and promising.

In summary, this work combines Transformer models and Retro* search algorithm to develop a leading-edge biosynthesis navigator (BioNavi-NP) for natural products, which can trace back to biologically-plausible building blocks and enumerate diverse biosynthetic pathways. BioNavi-NP outperforms other computational tools over many aspects of retro-biosynthesis analysis, thus it is promising for native biogenesis analysis, biosynthetic pathway reconstruction, and rational design.

**Methods**

As illustrated in Fig. 1b, all the single-step prediction were run with the ensemble of Transformer models[41], which were proved to be advantageous compared to the single model.[42, 43] Retro* searching algorithm[18] was used for multi-step planning with a cost (score) for ranking. More than 30,000 biochemical reactions were accessed from MetaCyc[23], KEGG[24] and MetaNetX[25], and the organic reactions whose components are similar to natural products were extracted from USPTO[22] data set for data augmentation. More details on the model and data preparation can be found in Supplementary Methods.

**Code availability**

The source code is available from the authors upon reasonable request.

**Author contributions**

R.W. designed and supervised the whole research. S. Z., T.Z. and Y. Y. contributed concept and implementation. All authors contributed to the interpretation of results. R.W., S.Z and T.Z wrote the manuscript. All authors reviewed and approved the final manuscript.

**Competing interests**

We declare competing interests. S.Z. and C.L. were employees of Galixir, and C.W.C. was advisor to Galixir during the course of this work.

**Acknowledgment**

This work was supported by the National Natural Science Foundation of China (21773313 and 61772566), National Key R&D Program of China (2020YFB020003), Guangdong Key Field R&D Plan (2019B020228001 and 2018B010109006), Guangzhou S&T Research Plan (202007030010), and Sun Yat-sen University (20ykzd13). We thank the Guangzhou and Shenzhen Supercomputer Center for providing computational source.

**Supplementary Information**

Supplementary methods, Figs. 1–4, and References.

**Keywords:** biosynthetic pathway, natural products, deep learning, retro-biosynthesis, transformer